\documentclass[aps,prb,floatfix,reprint,superscriptaddress]{revtex4-1}
\usepackage{graphicx}
\usepackage{bm,amsmath,amssymb,mathrsfs,dcolumn}
\usepackage[usenames]{color}
\usepackage{subfigure}
\usepackage{ulem}

\usepackage{color}

\usepackage{hyperref}
\hypersetup{colorlinks=true, linkcolor=blue, citecolor=red, urlcolor=blue}

\begin{document}
\title{Wiggling skyrmion propagation under parametric pumping}
\author{H. Y. Yuan}
\email[Electronic address: ]{yuanhy@sustc.edu.cn}
\affiliation{Department of Physics, Southern University of Science and Technology,
Shenzhen, 518055, Guangdong, China}

\author{X. S. Wang}
\affiliation{School of Microelectronics and Solid-State Electronics,
University of Electronic Science and Technology of China, Chengdu,
Sichuan 610054, China}
\affiliation{Physics Department, The Hong Kong University of Science and
Technology, Clear Water Bay, Kowloon, Hong Kong}

\author{Man-Hong Yung}
\email[Electronic address: ]{yung@sustc.edu.cn}
\affiliation{Institute for Quantum Science and Engineering and Department of Physics,
South University of Science and Technology of China, Shenzhen, 518055, China}
\affiliation{Shenzhen Key Laboratory of Quantum Science and Engineering, Shenzhen, 518055, China}

\author{X. R. Wang}
\email[Electronic address: ]{phxwan@ust.hk}
\affiliation{Physics Department, The Hong Kong University of Science and
Technology, Clear Water Bay, Kowloon, Hong Kong}
\affiliation{HKUST Shenzhen Research Institute, Shenzhen 518057, China}
\date{\today}

\begin{abstract}
We address the problem of how magnetic skyrmions can propagate along a guided direction
by parametric pumping. As evidenced by our micromagnetic simulations, skyrmions can
hardly be driven by either a static electric field or a static magnetic field alone.
Although the magnetic anisotropy can be modified by an electric field, parametric
pumping with an oscillating electric field can only excite the breathing modes.
On the other hand, a static magnetic field can break rotational symmetry through
the Zeeman interaction, but it cannot serve as an energy source for propelling a
skyrmion. Here we found that the combination of a perpendicularly oscillating electric
field and an in-plane static magnetic field can drive a skyrmion undergoing a wiggling
motion along a well defined trajectory. The most efficient driving occurs when the
frequency of the oscillating field is close to that of the breathing motion.
The physics is revealed in a generalized Thiele equation where a net spin current
excited by the parametric pumping can drive the skyrmion propagation through
angular momentum transfer. Compared with other alternative proposals, our results open
new possibilities for manipulating skyrmions in both metals and insulators with low-power
consumption. The oscillating skyrmion motion can also be a microwave generator for
future spintronic applications such as an nano-tool on a diamond Nitrogen-Vacancy center.
\end{abstract}

\maketitle

\section{Introduction}
Magnetic skyrmions are topological structures that were observed in a class of
magnetic materials with broken inversion symmetry \cite{Bogdanov2001,Rossler2006,
Muhlbauer2009,Yu2010,Yu2011,Woo2016,Yuan2016,Yuan2017}. In comparison with magnetic
bubbles \cite{Mal1979} and domain walls \cite{Yuan2015}, skyrmions are relatively small
(1-100 nm) \cite{Siemens2016}, and can be driven with a lower current density ($10^6~
\mathrm{A/m^2}$) \cite{Iwasaki2013}, making them ideal for being information carriers.
Recently, various methods have been proposed for controlling
skyrmion motion, including electric currents \cite{Iwasaki2013,Zhou2014,Woo2016,Kai2017},
spin waves \cite{Iwasaki2014,Schutte2014}, microwaves \cite{Wang2015,Yan2017}, and
temperature gradient\cite{Kong2013,Lin2014,Mochizuki2014}. In particular, skyrmions in
a metal driven by an electric current can move both parallel and transverse to
the current, known as the skyrmion Hall effect \cite{Kai2017,Jiang2017}.
However, a current does not work for insulating materials that may have lower damping,
lower power consumption, and better controllability. To manipulate skyrmions in insulators,
temperature gradient is proposed as a control knob through the spin transfer torque.
Unfortunately, similar to the magnonic spin transfer torque induced domain-wall motion,
\cite{Peng2011}, the effectiveness of thermal magnons remains a problem in practice.
Thus, finding new  control knobs for skyrmions is an interesting issue in spintronics.

Parametric pumping refers to a parameter cycling or oscillation that can result in a net
charge/spin transport. The system response to a parametric pumping may be strong (at
resonance) if the parameter cycling frequency matches with the system intrinsic frequency.
In recent years, using electric fields to manipulate magnetic states is a focus in
nanomagnetism \cite{Matsukura2015,Ohno2000,Ando2016,Dohi2016,Weisheit2007,Maruyama2009,
Lebeugle2009,Yang2016,Heron2011,Sch2011,Chiba2012,Franke2015}, because of its high
controllability and low energy consumption. Electric field can modify material parameters
such as exchange stiffness \cite{Ando2016,Dohi2016}, anisotropy coefficient
\cite{Weisheit2007,Maruyama2009,Lebeugle2009}, and even the strength of the
Dzyaloshinskii-Moriya interaction (DMI)~\cite{Dzy1957,Moriya1960,Yang2016}.
However, the behavior of a skyrmion subject to a parametric pumping remains a
largely-unexplored topic.

A perpendicularly-oscillating electric field (POEF) on a vertically-magnetized film can
periodically modify the magnetic anisotropy \cite{Weisheit2007,Maruyama2009,Lebeugle2009},
resulting in a parametric pumping. However, a skyrmion in such a film undergoes only a
breathing motion, instead of propagating along a well-defined direction.
In this paper, we show that a POEF together with an in-plane static magnetic field, which
breaks skyrmion rotational symmetry, can drive a skyrmion to move along a given direction.
The motion is attributed to the spin current that transfers its angular momentum to the
skyrmion wall and is associated with skyrmion breathing motion. The skyrmion velocity
reaches its maximum when the POEF frequency matches that of the breathing motion.
These results are numerically verified by micromagnetic simulations and are analytically
justified from the generalized Thiele equation.

\section{Model and methods}
\begin{figure}[t]
\includegraphics[width=0.95\columnwidth]{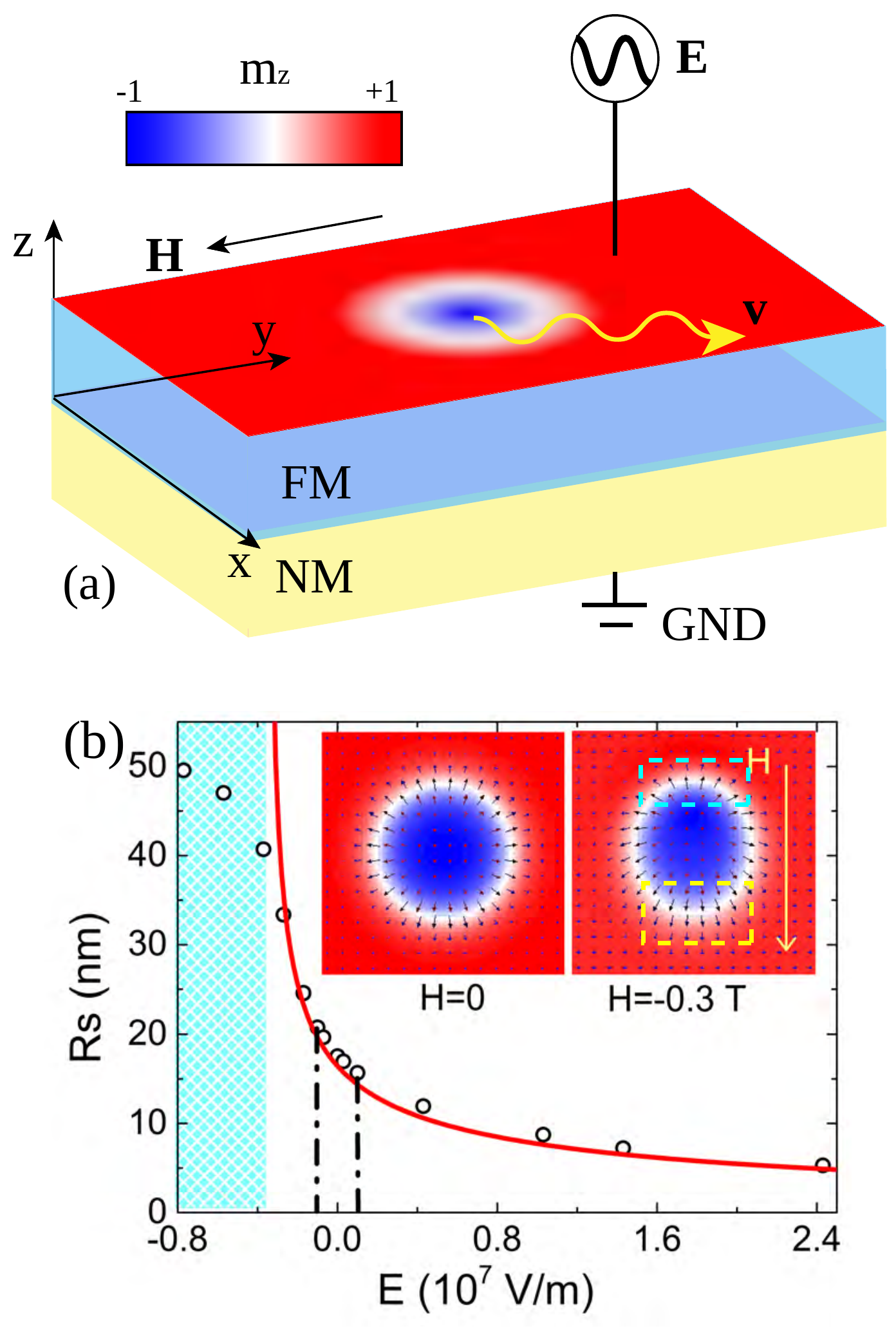}
\caption{(color online) (a) Schematic illustration of a magnetic layered structures.
The colors encode $m_z$ value with the color bar shown in the top-left panel.
A magnetic field along the y direction breaks the rotational symmetry of skyrmions with respect to
$z-$axis. An oscillating electric field is applied normally ($z$-direction) to the film.
The skyrmion moves wiggly in the $xy$-plane as illustrated by the yellow arrow.
(b) Skyrimion radius as a function of electric field in the absence of magnetic fields.
The blue shadowed region below $~-3.6 \times 10^6$ V/m is for the unstable skyrmions.
The red line is Eq. \eqref{radius}. The insets show the symmetric and asymmetric skyrmion structures
for $H=0$ (left) and -0.3 T (right), respectively. The dashed lines indicate the
field variation range within which $R_s$ varies from 20.8 nm to 15.7 nm.}
\label{fig1}
\end{figure}

We consider a perpendicularly-magnetized film with a skyrmion
in the center as shown in Fig. \ref{fig1}a. The skyrmion is stabilized by the competition
between exchange interaction, anisotropy and interface DMI \cite{Dzy1957,Moriya1960}
from the asymmetric interfaces of magnetic and non-magnetic layers.
The skyrmion dynamics is governed by the Landau-Lifshitz-Gilbert (LLG) equation,
\begin{equation}
\frac{\partial \mathbf{m}}{\partial t} =-\gamma\mathbf{m} \times \mathbf{H}_{\rm eff} +\alpha \mathbf m
\times \frac{\partial \mathbf{m}}{\partial t},
\label{llgt}
\end{equation}
where $\mathbf{m}$, $\gamma$, $\alpha$ are respectively the unit vector of the magnetization, gyromagnetic
ratio, and the Gilbert damping. $\mathbf{H}_{\rm eff}=2A \nabla^2\mathbf{m}+2K_um_z\mathbf{e}_z+\mathbf{H}_d
+H\mathbf{e}_y + \mathbf{H}_{\rm DM} + \mathbf{H}_{\rm so}$ is the effective field including the exchange
field, crystalline anisotropy field, dipolar field $\mathbf{H}_d$, external field $H\mathbf{e}_y$ along
the y-direction, DMI field $\mathbf{H}_{\rm DM}$ and spin-orbit field \cite{Up2015} $\mathbf{H}_{\rm so}$
due to the electric field. $A$ is the exchange stiffness and $K_u$ is the anisotropy coefficient.
The spin-orbit field is induced by the applied electric field through spin-orbit interaction and can
be divided into the damping-like components and field-like components \cite{Up2015}, i.e.
$\mathbf{H}_{\rm so} = \zeta_D E m_z \mathbf{m} \times \mathbf{e}_z + \zeta_F E m_z \mathbf{e}_z$,
here $E$ is the electric field while $\zeta_D$ and $\zeta_F$ are the torque conversion coefficients.
To investigate the skyrmion structure and its dynamics in an electric field, we use the  Mumax3 package
\cite{mumax3} to numerically solve the LLG equation. The film size is of $128 ~\mathrm{nm} \times 128 ~
\mathrm{nm} \times 0.4 ~\mathrm{nm}$, if it is not stated otherwise. The model parameters are
$A=10 \times 10^{-12} ~\mathrm{J/m}, D=0.003 ~\mathrm{J/m^2}, M_s=9.2 \times 10^5 ~\mathrm{A/m},
K_u=1.157 \times 10^6~\mathrm{J/m^3},\zeta_F=0.02~ \mathrm{J/(V\cdot m^2)}$ to mimic CoPd\cite{Up2015}.
The Gilbert damping varies from 0.02 to 0.2. We focus on the influence of field-like spin-orbit torque
on skyrmion dynamics and take $\zeta_D=0$ in the simulations.

\begin{figure}[t]
\includegraphics[width=0.9\columnwidth]{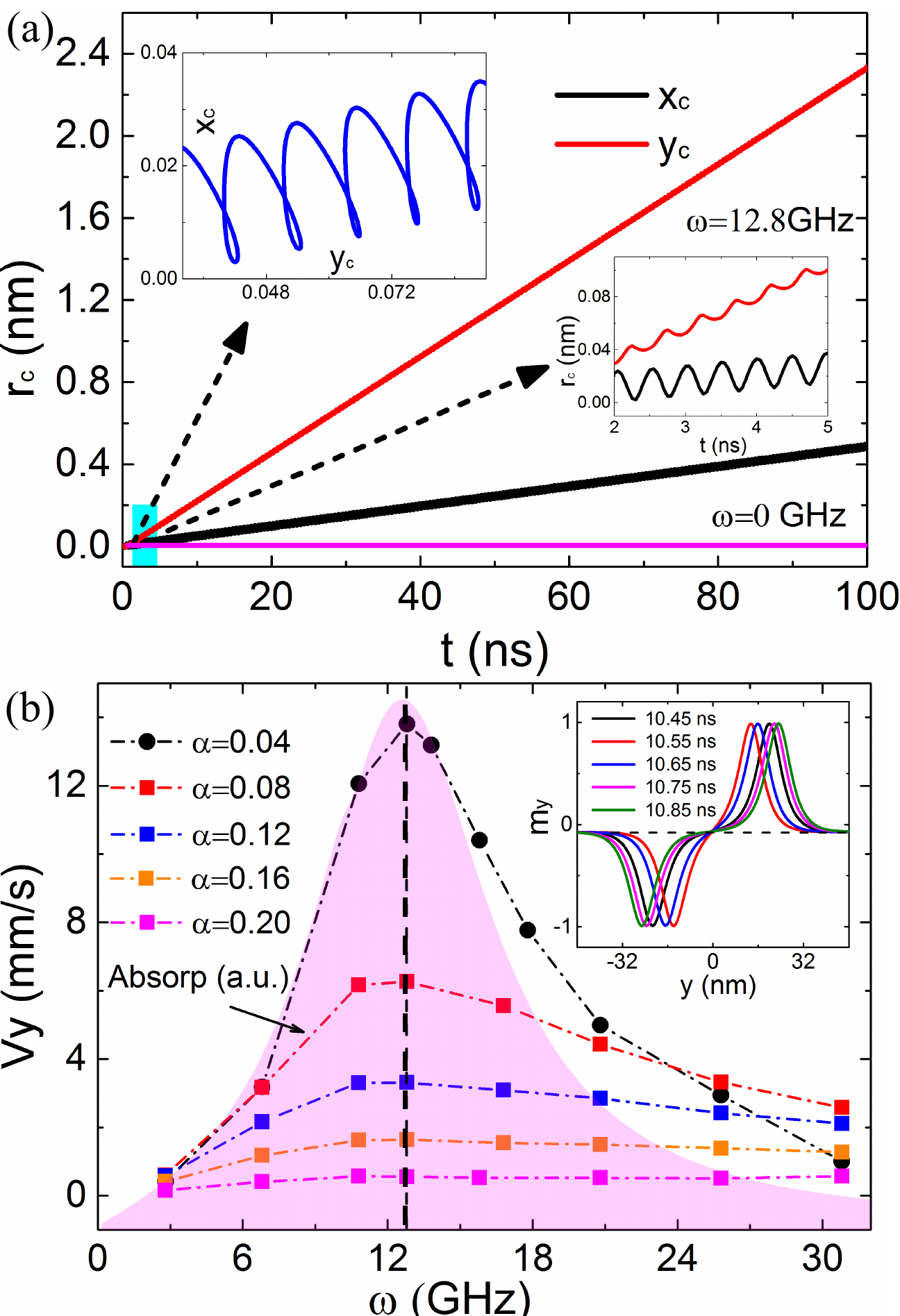}
\caption{(color online) (a) Evolution of skyrmion center $\mathbf{r}_c = (x_c,y_c)$ for $\omega=12.8$
GHz and $\omega =0$, respectively. The slope of the $x_c-t$ and $y_c-t$ curves give the skyrmion
velocity $v_x$ and $v_y$, respectively. The right inset is the zoom-in plots of $x_c$ (black) and $y_c$
(red) plots that show a wiggling skyrmion motion with a wiggling trajectory shown in the left inset.
(b) Average skyrmion velocity as a function of electric field frequency for $\alpha=0.04, 0.08,0.12,0.15$
and $0.20$, respectively. The pink shadow illustrates the absorption spectrum. The inset is the snapshots
of distribution of $m_y$ along $x=0$ at different times. The extremes locate the skyrmion wall centers.
The oscillations of wall centers with time demonstrate a breathing motion of the skyrmion under parametric
pumping.}
\label{fig2}
\end{figure}

\section{Results}
\subsection{Skyrmion structures}
Let us first look at the skyrmion structures under a static electric field ($E$).
Figure \ref{fig1}b shows that the skyrmion size $R_s$ decreases with the increase of electric field
with a typical skyrmion structure shown in the left inset for $E=0$ and $H=0$.
The skyrmion size can be described by \cite{xiansi2017}
\begin{equation}
R_s=\pi D \sqrt{\frac{A}{16AK_{\rm eff}^2 - \pi^2 D^2 K_{\rm eff}}},
\label{radius}
\end{equation}
where $K_{\rm eff}=K_u+\zeta_FE/2-\mu_0M_s^2/2$. Here the long-range dipolar interaction is approximated as
the shape anisotropy $\mu_0M_s^2/2$ along the $z-$axis, which is well justified for a magnetic thin film.
This is of the variational \cite{Rohart2013,Kravchuk2018,xiansi2017} result obtained by assuming the
skyrmion profile along radial direction as a $360^\circ$ domain wall with skyrmion size and skyrmion wall
width as two optimization parameters. The red solid line in Fig. \ref{fig1}b is Eq. \eqref{radius}
that describes well the simulation results (circles) for $E > -3.6 \times 10^6~\mathrm{V/m}$.
For electric fields smaller than that value, the skyrmion size (diameter$>80$ nm) is comparable with
the system size (128 nm) and the boundary effect becomes pronounced. In an infinite film, the skyrmion
should proliferate and becomes unstable \cite{Rohart2013} at the critical field $E_c=\pi^2 D^2/(8A\zeta_F)
-(2K_u-\mu_0M_s^2)/\zeta_F=-3.6 \times 10^6~ \mathrm{V/m}$ (blue shadowed region).
Under an in-plane field, the skyrmion deforms and elongates along the field direction as shown in the
right inset of Fig. \ref{fig1}b for $H=-0.3$ T. Here the top and bottom skyrmion walls become thinner
and thicker, respectively, to take the advantage of the Zeeman effect. The larger the in-plane field,
the larger the width difference between the top and the bottom skyrmion walls is (See the Appendix A).

\subsection{Skyrmion motion}
To describe the skyrmion motion of the asymmetric skyrmions under a harmonic POEF
of $\mathbf{E}=E_0 \sin (\omega t)\mathbf{e}_z$ and an in-plane magnetic field, we define the skyrmion
position as topological charge weighted center
\cite{Kong2013}:
$\mathbf{r}_c\equiv 1/(4\pi Q)\int \mathbf{m}\cdot(\partial_x \mathbf{m}\times\partial_y\mathbf{m})\mathbf{r}dS$
with $Q\equiv 1/(4\pi \int\mathbf{m}\cdot(\partial_x \mathbf{m}\times\partial_y\mathbf{m})dS$ being the skyrmion
number. Figure \ref{fig2}a shows the time dependence of skyrmion position for $\omega=0$ and 12.8 GHz, respectively.
When a static electric field is applied, i.e. $\omega=0$, the skyrmion does not move. When $\omega=12.8$ GHz,
the skyrmion shows a wiggling motion in both the $x-$ and the $y$-directions with typical trajectories
shown in the insets of Fig. \ref{fig2}a. Figure \ref{fig2}b is the field-frequency-dependence of the average
skyrmion velocity along the y-direction ($v_y$) for various damping coefficients ranging from 0.04 to 0.20.
$v_y$ is peaked around 12.8 GHz, almost independent of $\alpha$.

In order to check whether the peak is associated with the parametric resonance that occurs when the POEF
frequency matches with a skyrmion intrinsic frequency, we consider the dynamical susceptibility of the
system to a sine field of $E_z(t)=E_0 \sin {\omega t}/(\omega t)$, defined as
$\langle m_z(t) \rangle= \chi_{zz}(\omega) E_z(t) $, where $\langle m_z(t) \rangle$ is the average $m_z$.
The energy absorption of the system, proportional to $\mathrm{Im}(\chi (\omega))$ \cite{Yin2016},
is shown by the pink shadowed region in Fig. \ref{fig2}b.
The absorption peak is located around 13 GHz that coincides with the maximal skyrmion velocity.
The skyrmion response to the POEF of $\omega=12.8$ GHz is shown in the inset of Fig. \ref{fig2}b
that plots the snapshots of  $m_y$ along $x=0$. The positions with extreme $m_y$ values are the skyrmion
wall centers. The center positions oscillate back-and-forth with time. This shows clearly a strong breathing
motion of the skyrmion \cite{Mochizuki2012,Onose2012,Kim2014}. Thus, the velocity peak
corresponds to resonance of the POEF with the skyrmion breathing mode of 13 GHz. It should note a minor
peak around 10.4 GHz that will influences the skyrmion velocity for $\alpha <0.01$ (See the Appendix B).

\subsection{Generalized Thiele equation}
\begin{figure}
\includegraphics[width=0.9\columnwidth]{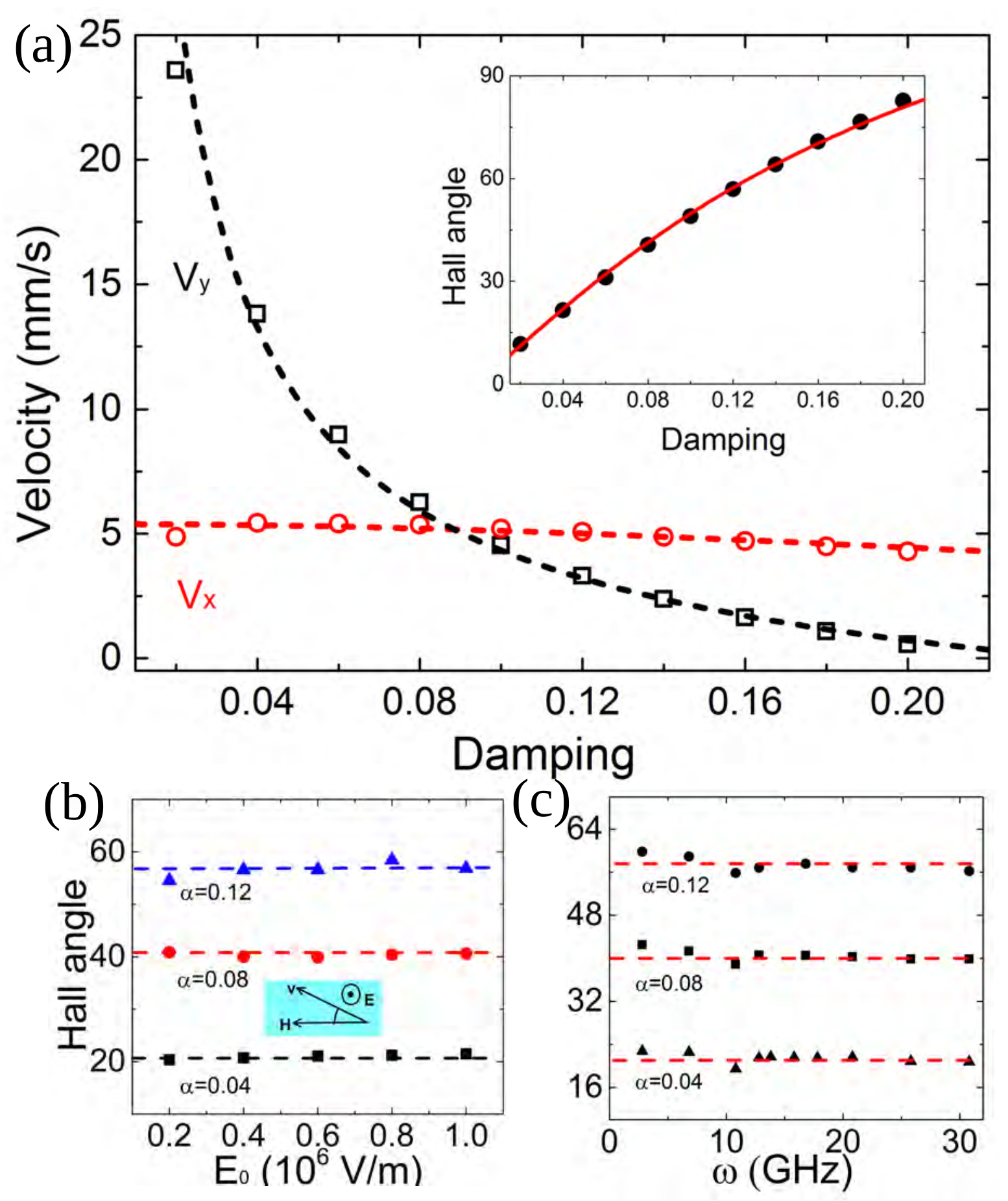}
\caption{(color online) (a) Average skyrmion velocity as a function of damping at $\omega=12.8$ GHz.
The symbols are simulation data, and the dashed lines are the solutions of the Thiele equation.
The inset shows the Hall angle defined as $\arctan(v_x/v_y)$ as a function of damping parameter.
The solid line is theoretical calculations.
(b) and (c) Electric field strength and field frequency dependence of Hall angle under different dampings.
The symbols are simulation data, and the horizontal dashed lines are used to guide eyes.}
\label{fig3}
\end{figure}

The wiggling motion of skyrmion center shown in Fig. \ref{fig2}a accompanies the skyrmion breathing.
The breathing motion emits spin waves (a magnon spin current) similar to the spin wave emission by
domain wall motion \cite{xiansi2012}. The emitted spin waves across the skyrmion wall, transfer the
angular momentum to a skyrmion and drive
the skyrmion to move, similar to spin transfer torque induced domain wall motion.
Because the rotational symmetry of the skyrmion is broken by the in-plane field, the magnon current
should have different components along the field direction ($+y$-direction) and the $x-$direction.
To understand the behavior, we consider the generalized Thiele equation \cite{Iwasaki2013} (See the Appendix C),
\begin{equation}
\mathbf{G} \times (\mathbf{v}-\mathbf{j}^{(m)}) + \mathbf{D} \cdot(\alpha \mathbf{v}-\beta \mathbf{j}^{(m)})=0
\label{thiele}
\end{equation}
where $\mathbf{G}=G\mathbf{e}_z=4\pi Q\mathbf{e}_z$ is the skyrmion gyrovector proportional to the skyrmion
number $Q$, and $D_{ij}=\int \partial_i \mathbf{m} \cdot \partial_j \mathbf{m}dS$ is the dissipation tensor.
$\mathbf{v}=(v_x,v_y)$ is average skyrmion velocity, and $\beta$ describes the mis-alignment of magnon
polarization and local magnetization that is zero here. $\mathbf{j}^{(m)}$ is the average magnon current.
The skyrmion velocity can be obtained from Eq. \eqref{thiele}
\begin{equation}
\begin{aligned}
v_x=\frac{ j^{(m)}_x - \alpha \kappa j^{(m)}_y}{1+\alpha^2 \kappa^2 },v_y=\frac{j^{(m)}_y
+ \alpha \kappa j^{(m)}_x}{1+\alpha^2 \kappa^2}\\
\end{aligned}
\label{veq}
\end{equation}
where $\kappa = D_{xx}/G$.
Figure \ref{fig3}a shows that $v_y$ decreases with the damping hyperbolically while $v_x$
is almost a constant, which suggests that the magnon current $j^{(m)}_y$ is inversely
proportional to $\alpha$ while $j^{(m)}_x$ is damping independent since
$\alpha \kappa \ll 1$ in Eq. (\ref{veq}), i.e. $j^{(m)}_y=C_y/\alpha, j^{(m)}_x=C_x$.
Using the parameters $C_x = 2\times 10^{-5}$, $C_y= 2.7\times 10^{-6}$, Eq. (\ref{veq}) can
indeed fit the numerical data (symbols) perfectly as shown in Fig. \ref{fig3}a.
Furthermore, the skyrmion Hall angle defined as $\mathrm{atan}(v_x/v_y)=\mathrm{atan}((j_x^{(m)}
-\alpha \kappa j_y^{(m)})/(j_y^{(m)} + \alpha \kappa j_x^{(m)}))$ is calculated and plotted as the red
line in the inset of of Fig. \ref{fig3}a. Again, it perfectly describes the numerical results (circles).
Interestingly, at given $\alpha$, the Hall angle is insensitive to both the amplitude and frequency of
electric field, as shown in Fig. \ref{fig3}b and c.

\begin{figure}
\includegraphics[width=0.9\columnwidth]{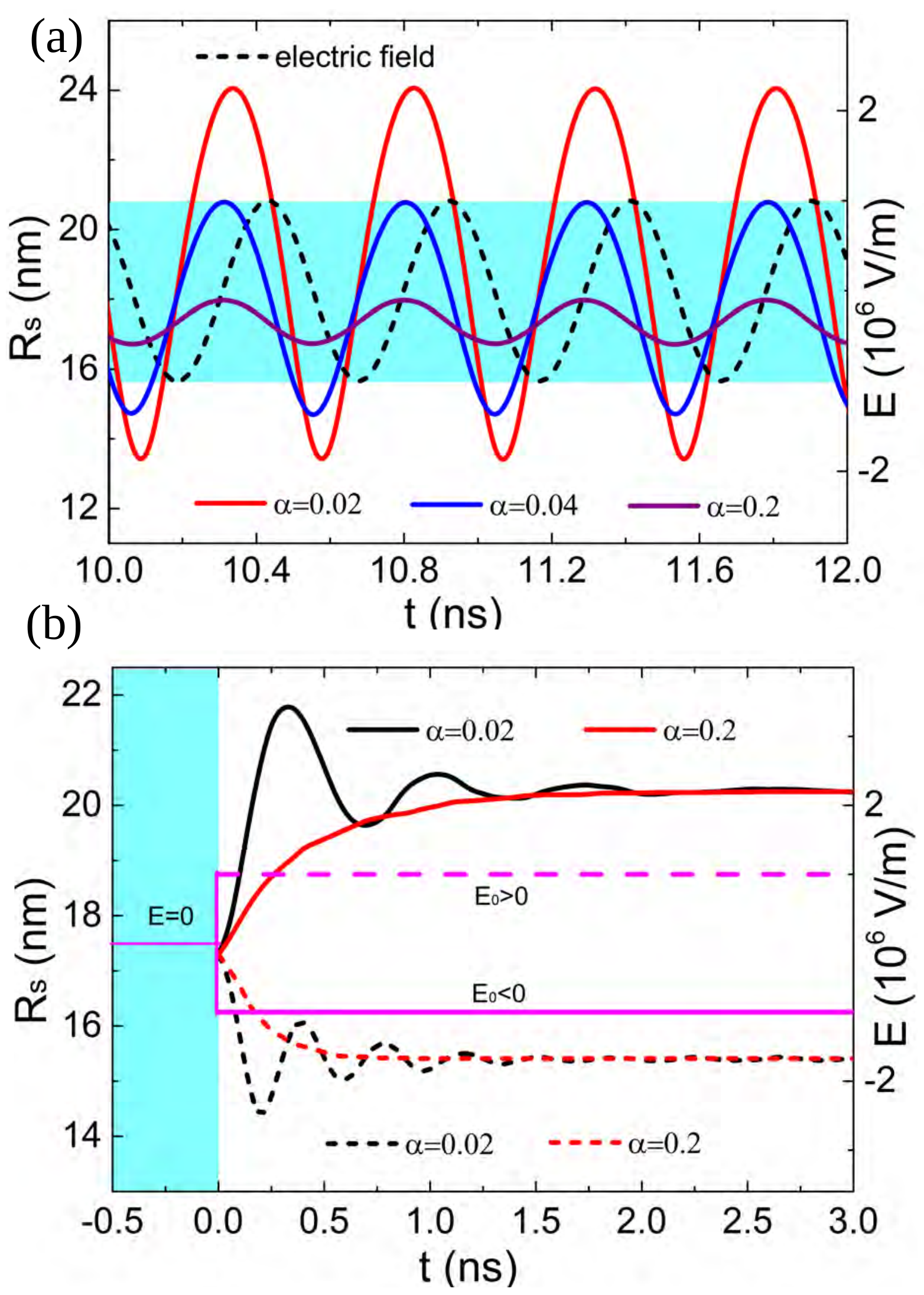}
\caption{(color online) (a) Skyrmion size as a function of time at $\omega=12.8$ GHz for
$\alpha=0.02$ (red line), 0.04 (blue line), and 0.20 (purple line), respectively.
The cyan rectangle is the region of size fluctuation by taking static field $E\in [E_0,-E_0]$.
The dashed line is the oscillation electric field.
(b) Time dependence of skyrmion size after a sudden quench of anisotropy by using a step electric
field switched on at $t=0$ns for $\alpha=0.02$ (black line) and and 0.2 (red line).
The solid and dashed lines are for $E_0<0$ (dashed horizontal line) and $E_0>0$ (solid horizontal line),
respectively.}
\label{fig4}
\end{figure}

\subsection{Skyrmion inflation and deflation}
The skyrmion size under an POEF oscillates periodically because the electric field modifies the
magnetic anisotropy. According to Eq. (\ref{radius}), the skyrmion size should vary in the range of
$R_{\min}=R_s(E=E_0)$ and $R_{\max}=R_s(E=-E_0)$ as shown in the cyan rectangle in Fig. \ref{fig4}a.
However, micromagnetic simulations show that the skyrmion size oscillates out of this range for
$\alpha \leq 0.04$ and falls into this range for $\alpha > 0.04$, as shown in Fig. \ref{fig4}a.
This indicates that skyrmion under parametric pumping has an inertia. The steady response of the
skyrmion size to an applied harmonic POEF (dashed lines) is shown in Fig. \ref{fig4}a for $\alpha=0.002$
(red), 0.04 (blue) and 0.2 (purple), respectively. They showed the typical breathing motion
(expansion-and-contraction). Evidently, the skyrmion size variation has a phase lag to its driving POEF.
The lagged phase increases with the damping, similar to a damped harmonic oscillator \cite{Fasano}.
It is under damped for a lower $\alpha$($\leq0.04$) so that the stored energy from POEF
will push the skyrmion to expand beyond its static size. It is over-damped for a larger $\alpha$.
The skyrmion motion lags behind the external pumping field so much that the skyrmion cannot reach its
maximal or minimal sizes corresponding to the minimal and maximal effective anisotropies.
To further substantiate the damping dependence of skyrmion size oscillation, we consider how the skyrmion
size responds to a sudden switching of a constant electric field. The results are shown in Fig. \ref{fig4}b.
The solid (dashed) lines are the evolution of skyrmion size $R_s$ to a constant electric field $E=1.0\times
10^6$V/m ($E=-1.0 \times 10^6$V/m) switched on at $t=0$ for $\alpha=0.02$ (black) and $\alpha=0.2$ (red),
respectively. The skyrmion size oscillates on its way to the equilibrium value for $\alpha = 0.02$ while
it takes a long time for $R_s$ to monotonically relax to its equilibrium value for $\alpha=0.2$.
Take $E<0$ as an example, the intermediate skyrmion size can be larger than the equilibrium value for
$\alpha=0.02$ while it is always smaller than the equilibrium value for $\alpha=0.2$.
For a fast oscillating electric field, the skyrmions size can be kept at intermediate values
periodically for small damping since the skyrmion cannot dissipate its energy timely.
This explains the observation of extraordinarily large/small skyrmion in Fig. \ref{fig4}a for small damping.

\subsection{Spin qubit manipulation}

The dipolar field outside the magnetic film will oscillate periodically accompanying the skyrmion size
oscillates under parametric pumping. Since the oscillation frequency of the dipolar field can reach GHz
level as shown in Fig. \ref{fig2}, the oscillating skyrmion can be a microwave generator useful for spin
qubit manipulation in quantum information science.

\begin{figure}
\centering
\includegraphics[width=0.48\textwidth]{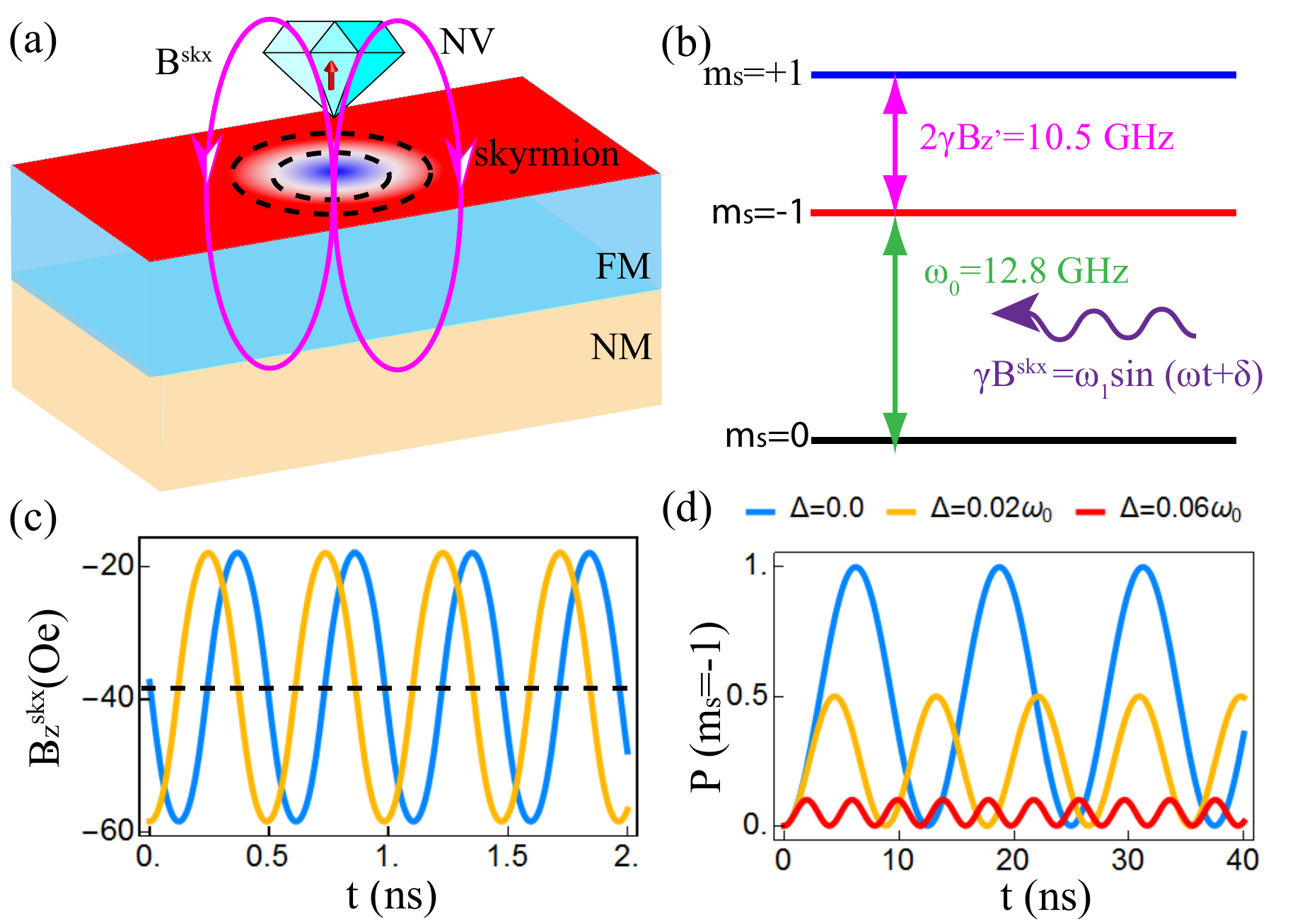}\\
\caption{(color online) (a) Schematic illustration of a FM/NM
bilayer with a nanodiamond placed on top of the FM layer. The spin qubit in the NV center of the
diamond interacts with the skyrmion via the dipolar field $\mathbf{B}^{\rm skx}$ generated by the
oscillating skyrmion (pink lines).
(b) Schematic illustration of the energy levels of the ground state of a NV center
under a Zeeman field $B_{z'}=298$ Oe. (c) The oscillation of the dipolar fields at $d=16$ nm with
a phase shift $\pi/2$.
(d) Population rate of the excited state ($m_s=-1$) as a function
of time for detuning $\Delta =0$ (blue line), $0.02~\omega_0$ (orange line),
and $0.06~\omega_0$ (red line), respectively.}
\label{fig5}
\end{figure}

Figure \ref{fig5}a shows a FM/NM bilayer with a nanodiamond placed on top of the FM layer. The spin qubit
inside the Nitrogen-Vacancy (NV) center of the diamond interacts with the skyrmion via the dipoalr interaction.
To see how the qubit in a NV center ($S=1$) responds to the oscillating dipolar field,
we recall the Hamiltonian of a NV center, \cite{Schirhagl2014}

\begin{equation}
H_{NV}=D_0(S_{z'})^2 + \gamma_{NV} S_{x'} (B^{skx}_{x'}+B_{x'}) + \gamma_{NV} S_{z'} B_{z'},
\end{equation}
where the $z'$-axis is chosen to align along one of the NV symmetry axes and the $x'$-axis
is along the magnetic film normal direction ($z$-axis).
$S_{x'}$, $S_{y'}$, and $S_{z'}$ are the spin-1 operators along the $x',y',z'$ directions,
$D_0/2\pi =2.87$ GHz is the zero-field splitting,
$\gamma_{NV}/2\pi = 2.8~ \mathrm{MHz/Oe}$ is the gyromagnetic ratio of nitrogen atom.
Here $B_{x'}$ is a bias field to cancel the direct part in $B^{skx}_{x'}$.
We have neglected the magnetic field generated by the oscillating electric field, which can
be estimated by solving the Maxwell equation as $L\omega E_0/(4c^2) \sim 10^{-9}$ T,
where $L=128$ nm is the film size, $\omega$ and $E_0$ are the frequency and amplitude
of the electric field, respectively, and $c$ is the speed of light.

Without external fields ($B_{z'}$), the ground state of a NV center is $m_s=0$
while the two excited states $m_s= \pm 1$ are degenerate in energy. With a static magnetic field,
the degeneracy of $m_s= \pm 1$ is broken, which results in a three level system of $m_s=0,-1,+1$, respectively,
as shown in Fig. \ref{fig5}b. Here we use a static magnetic field to tune the splittng of
$m_s=\pm 1$ such that the energy gap between $m_s=0$ and $m_s =-1$ is close to the frequency of
oscillating dipolar field generated by the breathing skyrmion ($\sim 12.8$ GHz for our parameters).
Moreover, we notice that energy shift induced by the dipolar field is much smaller than the
energy gap ($\omega_1 \ll \omega_0, \omega_1 \ll 2\gamma B_{z'}$), then the transition
between the energy levels $m_s=0$ and $m_s = -1$ dominates the absorption process and the
three level system can be reduced to a well-studied two level system.
By initializing the NV center to $m_s=0$ and turning on the breathing motion of skyrmions,
the population rate of $P (m_s=-1)$ will evolve from the initial state ($P=0$),
according to the well-known Rabi formula \cite{Sakura}
$P(m_s=-1) = |C(t)|^2 = \omega_1^2/\Omega^2 \sin^2 \left (\Omega t/2 \right)$,
where $\Omega = \sqrt{(\omega_0-\omega)^2 + \omega_1^2}$ is the Rabi frequency,
$\omega_0 = D_0-\gamma_{NV} B_{z'}$, $\omega_1= \gamma_{NV} B^{skx}_{x'}/\sqrt{2}$, $B^{skx}_{x'}$ is the oscillation
amplitude of the dipolar field, and $\Delta=\omega_0-\omega$ is
defined as the detuning of dipolar field from the resonant frequency,

Figure \ref{fig5}c shows a typical oscillation of the dipolar field generated by a breathing
skyrmion and Figure. \ref{fig5}d shows the oscillation of the population rate of $m_s=-1$
under this dipolar field for detuning $\Delta =0$ (blue line), $0.02~\omega_0$ (orange line),
and $0.06~\omega_0$ (red line), respectively.
At the resonant condition ($\Delta =0$), the population rate oscillates between
0 and 1 periodically with the frequency that is equal to the oscillation amplitude of
the dipolar field $\omega_1=252$ MHz, which is a typical Rabi signal. As the detuning increases
to $\Delta=0.02~\omega_0$, the maximum population of the excited state is near 0.5.
As the detuning increases further, the population of the excited state keeps decreasing
and finally approaches zero.

\section{Discussions and Conclusions}
In conclusion, the combination of parametric pumping by a POEF and an in-plane static
magnetic field can drive skyrmions undergoing a wiggling motion.
The skyrmion velocity reaches its maximum value when the POEF frequency matches
with the skyrmion breathing frequency. Our results show a promising avenue
for manipulating skyrmions motion in both metallic and insulating magnetic materials.
Moreover, the role of in-plane field may be replaced by the exchange bias field in a
FM/Antiferromagnet bilayer such that all electric control of skyrmion dynamics can be realized.

Remarkably, temperature-gradient driven skyrmions exhibit a similar damping dependence of the
skyrmion velocity as those reported here by parametric pumping. Specifically, the longitudinal
(field-direction) velocity quickly decreases with damping while the transverse velocity is
insensitive to the damping \cite{Kong2013}. The skyrmion velocity under the two driven
forces are at the same order of cm/s \cite{Kong2013,Lin2014}. These coincidence may be
attributed to the fact that both the electric field and thermal driven skyrmion motion originate
from non-uniform magnon flow. Moreover, the skyrmion Hall angle induced by parametric pumping
is insensitive to both pumping frequency and pumping amplitude as shown in Fig. \ref{fig3}b and c.
This feature is desirable in manipulating skyrmion trajectory in practice.

Although our simulations focus on the N\'{e}el skyrmions, the physics should be applicable
to Bloch skyrmions (See the Appendix D). Moreover, parametric pumping can also be realized through the
cycling of the exchange stiffness and DMI strength besides of the anisotropy studied here.
One should expect similar behavior of the skyrmion motion as that in Fig. \ref{fig1}a (See the Appendix E)
when other parameter cycling is used. In this sense, parametric pumping is a universal control knob
for skyrmion motion. As a comparison, the combined interaction of microwave field and an in-plane field
could drive a skyrmion to move in a straight line without any wiggling \cite{Wang2015}. The Hall angle dramatically depends
on the in-plane field as well as the microwave frequency, which is very different from our observation shown
in Fig. \ref{fig3}bc. The anomalous skyrmion size oscillation shown in Fig. \ref{fig4} was not found in
those publications.

\section*{Acknowledgments}
HYY acknowledges the help communication with Weiwei Wang.
This work was financially supported by National Natural
Science Foundation of China (Grants No. 61704071),
Natural Science Foundation of Guangdong Province (2017B030308003) and
the Guangdong Innovative and Entrepreneurial Research Team Program (No. 2016ZT06D348),
and the Science Technology and Innovation Commission of Shenzhen Municipality
(ZDSYS20170303165926217, JCYJ20170412152620376).
XRW was supported by the NSFC Grant (No. 11774296) as well
as Hong Kong RGC Grants (Nos. 16301518 and 16301816).

\section*{Appendix A: In-plane field dependence of skyrmion profile}
\begin{figure*}
\includegraphics[width=1.8\columnwidth]{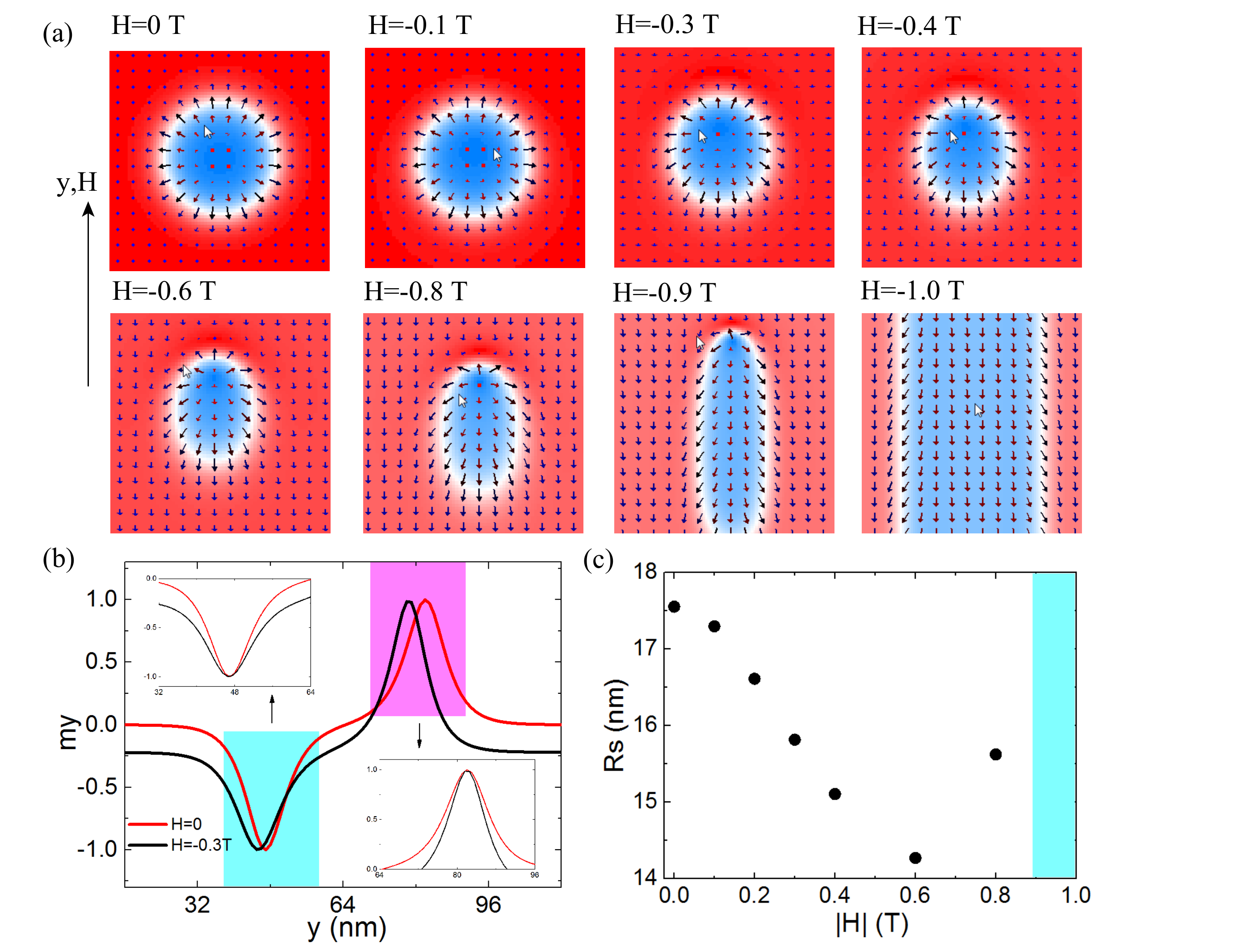}
\caption{(color online) (a) Spin configurations of the system as in-plane field decreases from 0 to -1.0 T.
(b) $m_y\sim y$ for spins along the line $x=64$ nm under $H=0$ (red line) and $H=-0.3$ T (black line).
The insets show the comparison of the skyrmion wall width
with and without applied fields for cyan region and pink region, respectively.
(c) Skyrmion size as a function of in-plane field strength. The skyrmion becomes unstable for $H>0.9$ T (cyan region). Here the skyrmion size ($R$) is defined by first counting the number of spins with $m_z>0$ ($N$), and then solving $R$
from the algebraic equation $\pi R^2 =Nd^2$, where $d$ is the mesh size.}
\label{sfig1}
\end{figure*}

Figure \ref{sfig1}a shows the spin configurations as the in-plane field decreases from 0 T to -1.0 T. For $|H|< 0.9$ T, the skyrmion
deforms more and more severely with the increase of fields and finally becomes unstable for $|H|\geq 0.9$ T. The skyrmion size
first decreases and then increases slightly with the field as shown in Fig. \ref{sfig1}c. To see the asymmetric deformation clearly,
a typical spin distribution in the y direction is plotted in Fig. \ref{sfig1}b. Here the skyrmion wall with spins parallel to the field
expands (cyan region) while the skyrmion wall with anti-parallel orientations with the fields shrinks (pink region).

\section*{Appendix B: Generalized Thiele equation}
To derive the generalized Thiele equation that describes the drift motion of skyrmion center,
we start from the Landau-Lifshitz-Gilbert (LLG) equation
that governs the dynamic precessions of the spins inside a skyrmion, i.e.,
\begin{equation}
\frac{\partial \mathbf{m}}{\partial t} =-\gamma\mathbf{m} \times \mathbf{H}_{\rm eff} +\alpha \mathbf m
\times \frac{\partial \mathbf{m}}{\partial t}.
\label{llgt}
\end{equation}
where $\mathbf{m}$ is the normalized magnetization, $\gamma$ is gyromagnetic ratio,
$\alpha$ is Gilbert damping that represents the energy dissipation rate of the
system, $\mathbf{H}_{\rm eff}$ is the effective field acting on the magnetization.
To distinguish the drifting of skyrmion position and the oscillation of skyrmion size, we decompose
the magnetization motion into a slow motion mode ($\mathbf{m}_s$) and a fast motion mode
($\mathbf{m}_{\rm f}$), i.e. $\mathbf{m}=(1-\mathbf{m}_f^2)\mathbf{m}_s + \mathbf{m}_{\rm f}$, where
the slow mode represents the equilibrium configuration evolution of the skyrmions while the fast mode
refers to the spin wave excitation around the equilibrium configuration of skyrmion.
Substituting the decomposition back into the LLG equation (\ref{llgt}) and taking a long time (many
oscillation periods of the fast mode) average, the dynamic equation of the slow mode can be written
as \cite{Kong2013}
\begin{equation}
\frac{\partial \mathbf{m}_s}{\partial t} =-\gamma \langle \mathbf{m}_{\rm f} \times \mathbf{H}_{\rm f}
\rangle +\alpha \mathbf m_s\times \frac{\partial \mathbf{m}_s}{\partial t}.
\label{llgs}
\end{equation}
where $\mathbf{H}_\mathrm{f}$ is the revised effective field $\mathbf{H}_{\rm eff}(\mathbf{m})$ with
$\mathbf{m}$ replaced by $\mathbf{m}_\mathrm{f}$. Since $\mathbf{m}_\mathrm{f}$ represents the magnon
excitation around equilibrium configuration, $\langle \mathbf{m}_\mathrm{f} \times \mathbf{H}_{\rm f}\rangle$
can be interpreted as magnon flow in the system \cite{Kong2013}.
For steady skyrmion motion, the translational symmetry of the skyrmion structures gives $\mathbf{m}_s(\mathbf{r}_c) =\mathbf{m}_s(\mathbf{r}_c-\mathbf{v}t)$, where $\mathbf{v}$ is skyrmion velocity and $\mathbf{r}_c$ is the short time average
 of skyrmion position, such that $\partial_t \mathbf{m}_s = -\mathbf{v} \cdot \nabla \mathbf{m}_s$. Performing the operation,
$\int \mathbf{m}_s \cdot (\nabla \mathbf{m}_s \times$ Eq. (\ref{llgs})), we obtain the generalized Thiele equation,
\begin{equation}
\mathbf{F}+\mathbf{G} \times \mathbf{v} + \alpha \mathbf{D} \cdot \mathbf{v}=0,
\label{thiele}
\end{equation}
where $\mathbf{G}=4\pi Q\mathbf{e}_z$ is the gyrovector of the skyrmion with $Q=\pm 1$ the topological charge of skyrmion,
$D_{ij}=\int \partial_i \mathbf{m} \cdot \partial_j \mathbf{m}dS$ is dissipation tensor, and
$\mathbf{F}=-\int  \langle \mathbf{m}_{\rm f} \times \mathbf{H}_{\rm f}\rangle \cdot ( \mathbf{m}_s \times \nabla \mathbf{m}_s) dxdy$
is the driven force coming from the magnon flow. The skyrmion velocity can be solved as
\begin{equation}
v_i=-\frac{\alpha D F_i + 4\pi Q \epsilon_{3ij}F_j}{(4\pi Q)^2+\alpha^2D^2}
\end{equation}
where $i,j=1,2,3$ corresponds to $x,y,z$ coordinates. Alternatively, the dynamic equation Eq. (\ref{thiele}) can be written as

\begin{equation}
\mathbf{G} \times (\mathbf{v}-\mathbf{j}^{(m)}) + \alpha \mathbf{D} \cdot \mathbf{v}=0,
\label{thiele2}
\end{equation}
where the magnon current $\mathbf{j}^{(m)}$ is defined as $\mathbf{F}=-\mathbf{G} \times \mathbf{j}^{(m)}$.
This form of Thiele equation is adopted in the main text.

For an arbitrary magnetic structure ($\mathbf{m}_s$), the spin wave excitation can be written as
$\mathbf{m}_{\rm f}=(m_\theta \mathbf{e}_\theta + m_\varphi \mathbf{e}_\varphi)\mathbf{e}^{i\omega t}$, where $\mathbf{e}_\varphi$
and $\mathbf{e}_\theta$ are determined by the local magnetization direction $\mathbf{e}_r \equiv \mathbf{m}_s$.
Then the effective fields due to exchange interaction, anisotropy term, Dzyaloshinskii-Moriya
interaction term (DMI) and Zeeman term read,

\begin{widetext}
\begin{equation}
\begin{aligned}
&\mathbf{H}_{\rm f,ex}=2A\nabla^2 \mathbf{m}_f =2A\left [ -m_\theta \nabla^2 \theta - \nabla m_\theta \cdot \nabla \theta \right ]\mathbf{e}_r +2A\left [ \nabla^2 m_\theta -  m_\theta (\nabla \theta)^2 \right ]\mathbf{e}_\theta +2A\nabla^2 m_\varphi \mathbf{e}_\varphi\\
&\mathbf{H}_{\rm f,an}=2Km_{f,z}\mathbf{e}_z=-2Km_\theta \sin \theta (\cos \theta \mathbf{e}_r - \sin \theta \mathbf{e}_\theta)\\
&\mathbf{H}_{\rm f,DM}=D[\nabla \cdot \mathbf{m} \mathbf{e}_z - \nabla m_z]=D\partial_y (m_\theta \mathbf{e}_\theta + m_\varphi \mathbf{e}_\varphi)_ye_z- \partial_y (m_\theta \mathbf{e}_\theta + m_\varphi \mathbf{e}_\varphi)_z \mathbf{e}_y\\
&\mathbf{H}_{\rm f,ze}=0\\
\end{aligned}
\end{equation}
\end{widetext}
where we have assumed that the azimuthal angle $\phi$ is space independent, i.e. $\partial_y \varphi=0$.
Then the exchange contribution to the driven force can be written as
\begin{widetext}
\begin{equation}
\begin{aligned}
F_{\rm ex,y} &= -\int \langle \mathbf{m}_{\rm f}\times \mathbf{H}_{\rm f,ex}\rangle \cdot (\mathbf{m}_s \times \partial_y \mathbf{m}_s) = -2A\int \partial_y \theta m_\theta \left[ m_\theta \nabla^2 \theta + \nabla m_\theta \cdot \nabla \theta \right ]dy
\end{aligned}
\end{equation}
\end{widetext}
where
\begin{widetext}
\begin{equation}
\begin{aligned}
\mathbf{m}_s \times \partial_y \mathbf{m}_s&=\partial_y \theta (-\sin \varphi e_x + \cos \varphi e_y)\\
\langle \mathbf{m}_{\rm f}\times \mathbf{H}_{\rm f,ex}\rangle_x &= 2A\left [ \sin \varphi m_\theta + \cos \theta \cos \varphi m_\varphi \right ]\left [ -m_\theta \nabla^2 \theta - \nabla m_\theta \cdot \nabla \theta \right ] \\
&+2A \sin \theta \cos \theta \varphi m_\theta \nabla^2 m_\varphi - 2A \sin \theta \cos \varphi m_\varphi [\nabla^2 m_\theta - m_\theta (\nabla \theta)^2]\\
\langle \mathbf{m}_{\rm f}\times \mathbf{H}_{\rm f,ex}\rangle_y &=  2A\left [ -\cos \varphi m_\theta + \cos \theta \sin \varphi m_\varphi \right ]\left [ -m_\theta \nabla^2 \theta - \nabla m_\theta \cdot \nabla \theta \right ] \\
&+2A \sin \theta \sin \theta \varphi m_\theta \nabla^2 m_\varphi - 2A \sin \theta \sin \varphi m_\varphi [\nabla^2 m_\theta - m_\theta (\nabla \theta)^2]\\
\end{aligned}
\end{equation}
\end{widetext}
Similarly, we can derive the contribution to $F_y$ from the anisotropy, DM interaction and Zeeman field as
\begin{equation}
\begin{aligned}
&F_{\rm an,y}=-2K\int \partial_y \theta \sin \theta \cos \theta m_\theta^2 dy\\
&F_{\rm DM,y}=\frac{D}{2} \int \partial_y \theta \partial_y m_\theta^2 dy\\
&F_{\rm ze,y}=0.\\
\end{aligned}
\end{equation}

\begin{figure*}
\includegraphics[width=1.8\columnwidth]{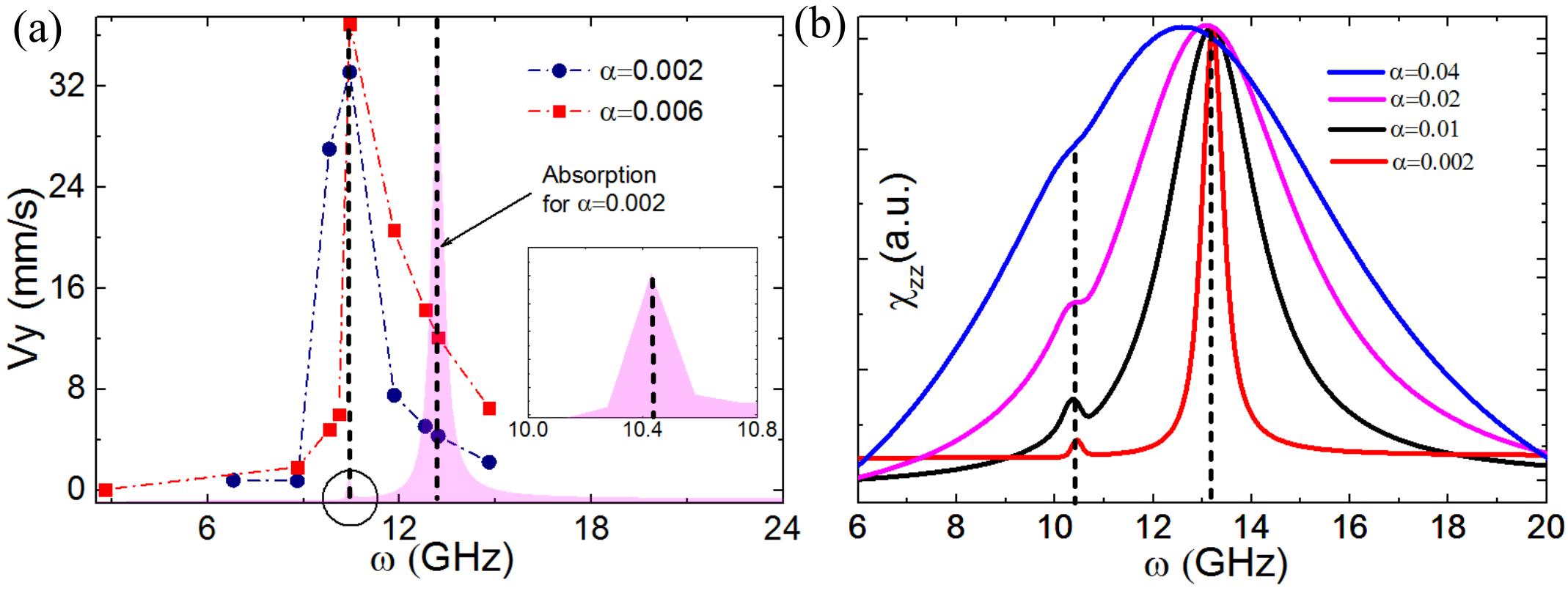}
\caption{(color online) (a) Skyrmion velocity as a function of electric field frequency for
$\alpha=0.002$ (blue dots) and 0.006 (red squares), respectively. (b) Absorption spectrum of
the magnetic system for $\alpha=0.04$ (blue line), 0.02 (pink color), 0.01 (black line), 0.002 (red line),
respectively. The dashed lines refer to the positions of resonance peaks}
\label{sfig2}
\end{figure*}

\begin{figure*}
\includegraphics[width=1.8\columnwidth]{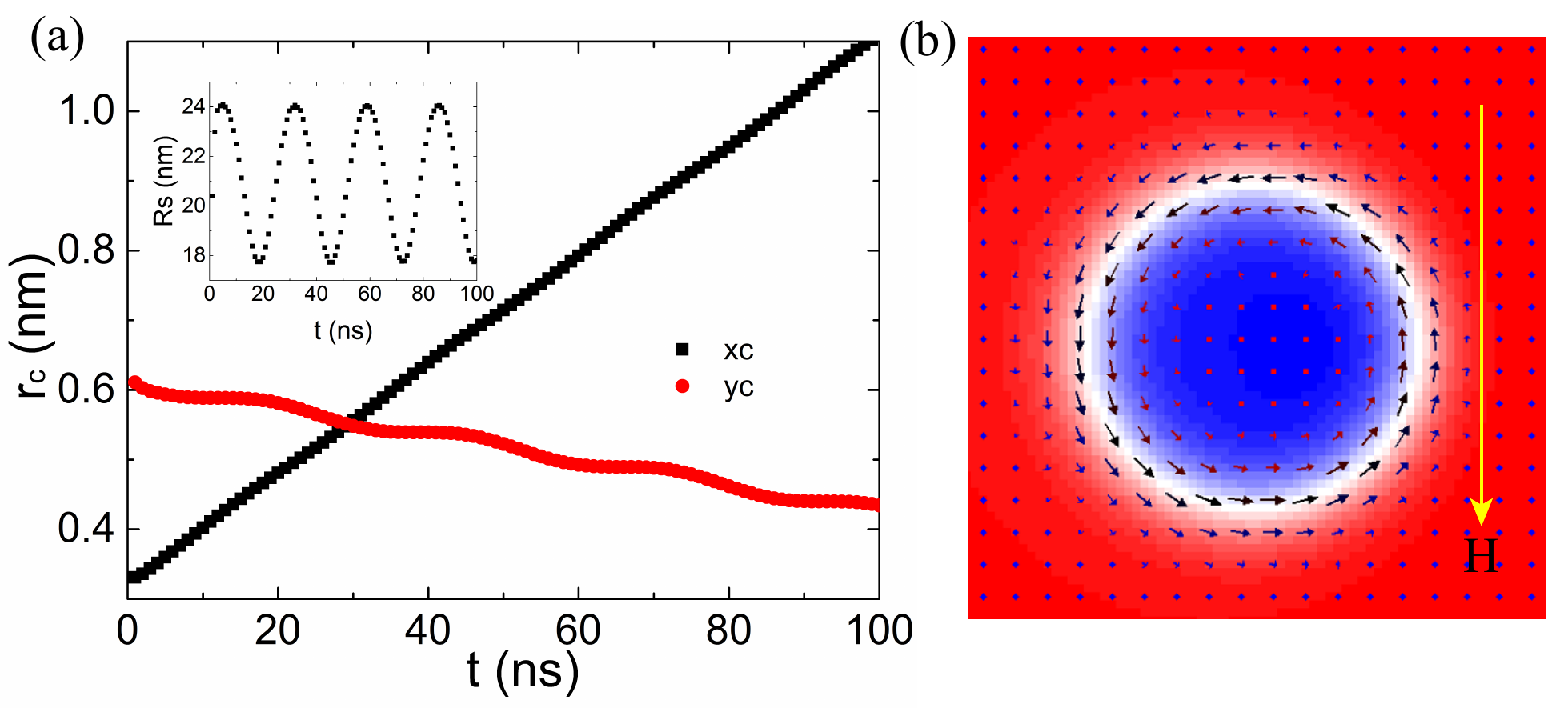}
\caption{(color online) (a) Bloch skyrmion position as a function of time by periodically tuning the
anisotropy constant $K=K_u + \Delta K \sin (\omega t)$. The parameters are $A=10^{-11}$ J/m, The bulk DMI strength
$D =3~ \mathrm{mJ/m^2}$,
$Ku=11.57 \times 10^5~ \mathrm{J/m^3}$,$\Delta K=0.1 \times 10^5~ \mathrm{J/m^3}$, $\alpha=0.02$, $\omega=12.8$ GHz, $H=-0.1$ T. The sampling rate is 1 frame/ns.
The inset shows the skyrmion radius as a function of time. (b) A static skyrmion profile. $H=-0.1$ T,
the bulk DMI strength $D =3~ \mathrm{mJ/m^2}$.}
\label{sfig3}
\end{figure*}

\begin{figure*}
\includegraphics[width=1.8\columnwidth]{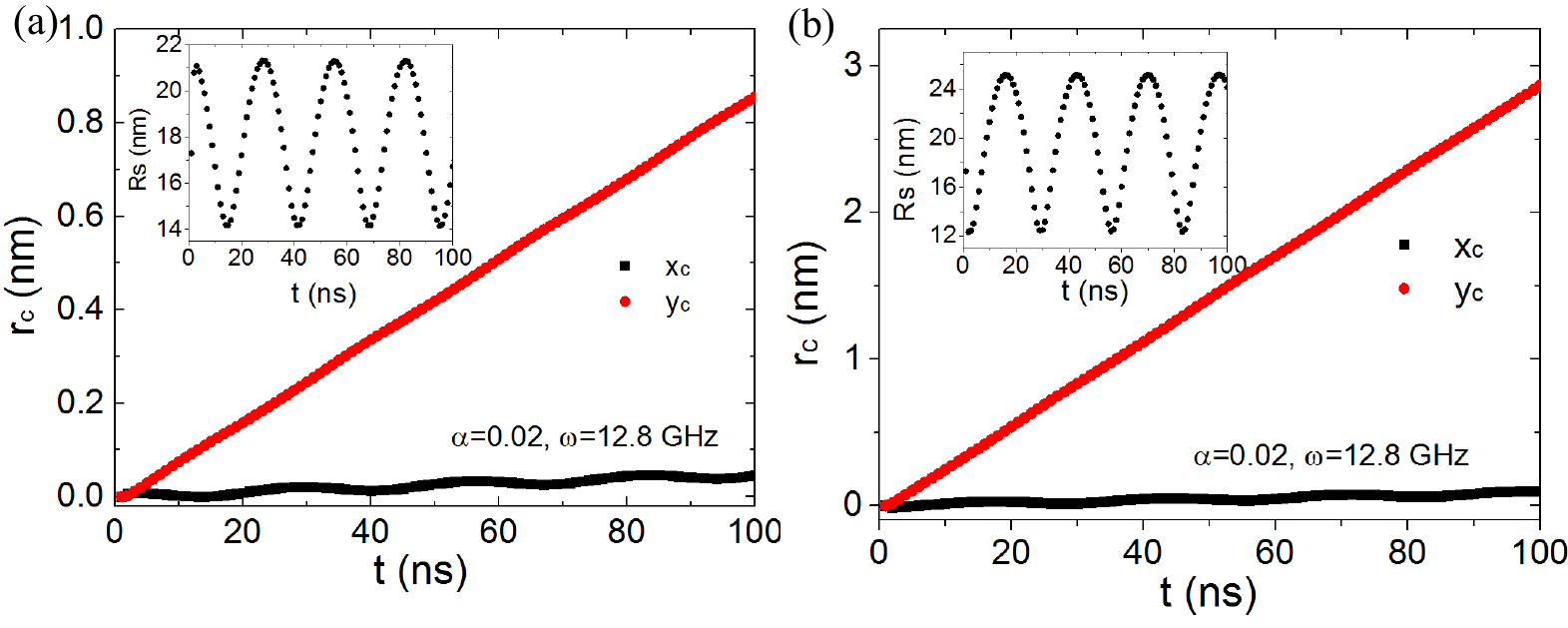}
\caption{(color online) Skyrmion position as a function of time by periodically tuning the exchange constant
$A=A_0 + 0.01A_0 \sin (\omega t)$ (a) and $D=D_0 + 0.01D_0 \sin (\omega t)$ (b), respectively. The inset shows
the skyrmion radius as a function of time. The parameters are $A_0=10^{-11}$ J/m, $D_0 =3~ \mathrm{mJ/m^2}$,
$K_u=11.57 \times 10^5~ \mathrm{J/m^3}$. $H=0.1$ T. The sampling rate is 1 frame/ns.}
\label{sfig4}
\end{figure*}

It has been shown that the spin orientation in the radial direction of a skyrmion can be well
described by a 360 domain wall \cite{xiansi2017} in the form,
\begin{equation}
\theta (r) = 2\arctan \left [ \frac{\sinh(R/\Delta)}{\sinh (r/\Delta)}\right ],
\end{equation}
where $R$ is skyrmion radius and $w$ is skyrmion wall width. Given $R\gg w$, this profile can be approximated as
\begin{equation}
\theta (r) = -2\arctan \left [ \exp\frac{r-R}{\Delta} \right ],
\end{equation}
which is the Walker profile for a $180^\circ$ domain wall, which will be used to further
simplify the driven force. The driven force in the y-direction becomes
\begin{equation}
\begin{aligned}
&F_{\rm ex,y}=-\frac{2A}{\Delta^3}\int \sin^2 \theta m_\theta (\cos \theta m_\theta + \Delta \partial_y m_\theta)dy\\
&F_{\rm an,y}=-\frac{2K}{\Delta}\int \sin^2 \theta \cos \theta m_\theta^2 dy\\
&F_{\rm DM,y}=\frac{D}{2\Delta} \int \sin \theta \partial_y m_\theta^2 dy\\
&F_{\rm ze,y}=0\\
\end{aligned}
\end{equation}
where we have used the relations $\partial_y \theta = \sin \theta/\Delta$,
$\partial_{yy} \theta = \sin \theta \cos \theta /\Delta^2$ that is true for Walker profile of a magnetic structure.

In summary, the total force is
\begin{widetext}
\begin{equation}
F_y=-\frac{2A}{\Delta^3}\int \sin^2 \theta m_\theta (\cos \theta m_\theta + \Delta \partial_y m_\theta)dy
-\frac{2K}{\Delta}\int \sin^2 \theta \cos \theta m_\theta^2 dy+\frac{D}{2\Delta} \int \sin \theta \partial_y m_\theta^2 dy
\end{equation}
\end{widetext}

For a skyrmion with rotational symmetry, the spin wave excitation of the symmetric skyrmion wall
($H=0$ T in Fig. \ref{sfig1}a) is also symmetric, hence $F_y=0$. For an asymmetric skyrmion, the
spin wave excitation becomes asymmetric, where the narrower skyrmion wall (smaller $\Delta$) emit
spin waves more intensively than the wider skyrmion wall (larger $\Delta$) as shown in Fig. \ref{sfig1}b.
Hence a net driven force in the asymmetric direction ($y$) become non-zero.
Moreover, due to the skyrmion Hall effect, the motion of skyrmion along the $y$ direction will induce a
skyrmion motion along the $x$ direction and consequently deforms the skyrmion in the $x$ direction.
As a result, a finite $F_x$ exist.

%

\section*{Appendix C: Skyrmion velocity in the low damping regime}
Figure \ref{sfig2}a shows the skyrmion velocity as a function of field frequency for $\alpha=0.002$
(blue dots) and 0.006 (red squares), respectively. The position of maximum velocity shifts to 10.4 GHz,
where a small absorption peak is identified. This suggests that the new mode makes significant contribution
to the skyrmion velocity in the low damping regime. In larger damping regime, the role of this mode is dominated
by the major mode around 13 GHz as shown in Fig. \ref{sfig2}b.

\section*{Appendix D: Bloch skyrmion propagation driven by parametric pumping}
In this section, we show that Bloch skyrmion can also be driven to move under the periodical oscillation
of magnetic parameters. Figure \ref{sfig3}a shows the skyrmion position as a function of time for a Bloch skyrmion
under the influence of the in-plane field $H_y=-0.1$ T and the periodic pumping of magnetic anisotropy. The inset shows
the oscillation of skyrmion radius. Figure \ref{sfig3}b shows the static profile of an asymmetric Bloch skyrmion.

\section*{Appendix E: Skyrmion propagation by oscillating exchange stiffness and DMI strength}
In this section, we show two examples of moving skyrmions by periodically changing
exchange stiffness and DMI. As shown in Fig. \ref{sfig4}, the skyrmion obtains
a finite speed of 8.5 mm/s and 30 mm/s by periodically tuning the exchange stiffness
and DMI strength by only $1\%$ percent while the skyrmion size
oscillates during the propagation, which is similar to the case by tuning the anisotropy.

\newpage

\begin{thebibliography}{}
\bibitem{Bogdanov2001} A. N. Bogdanov and U. K. R\"{o}{\ss}ler, Phys. Rev. Lett.
\textbf{87}, 037203 (2001).

\bibitem{Rossler2006}U. K. R\"{o}{\ss}ler, A. N. Bogdanov, and
C. Pfleiderer, Nature {\bf 442}, 797 (2006).

\bibitem{Muhlbauer2009} S. M\"{u}hlbauer, B. Binz, F. Jonietz, C. Pfleiderer, A. Rosch,
A. Neubauer, R. Georgii, P. B\"{o}ni1, Science {\bf 323}, 915 (2009).

\bibitem{Yu2010} X. Z. Yu, Y. Onose,, N. Kanazawa, J. H. Park, J. H. Han,
Y. Matsui, N. Nagaosa, and Y. Tokura, Nature \textbf{465}, 901 (2010).

\bibitem{Yu2011} X. Z. Yu, N. Kanazawa, Y. Onose, K. Kimoto, W. Z. Zhang, S. Ishiwata, Y. Matsui, and Y. Tokura
 Nat. Mater. \textbf{10}, 106 (2011).

\bibitem{Woo2016} S. Woo, K. Litzius, B. Kr\"{u}ger, M.-Y. Im, L. Caretta, K. Richter,
M. Mann, A. Krone, R. M. Reeve, M. Weigand, P. Agrawal,
I. Lemesh, M.-A. Mawass, P. Fischer, M. Kl\"{a}ui,
and G. S. D. Beach, Nat. Mater. \textbf{15}, 501 (2016).

\bibitem{Yuan2016} H. Y. Yuan and X. R. Wang, Sci. Rep. \textbf{6}, 22638 (2016).

\bibitem{Yuan2017} H. Y. Yuan, O. Gomonay, and Mathias Kl\"{a}ui, Phys. Rev. B
\textbf{96}, 134415 (2017).

\bibitem{Mal1979} A. P. Malozemoff and J. C. Slonczewski, \textit{Magnetic
domain walls in bubble materials} (Academic Press, 1979).


\bibitem{Yuan2015} H. Y. Yuan and X. R. Wang, Phys. Rev. B \textbf{92}, 054419 (2015).

\bibitem{Siemens2016} A. Siemens, Y. Zhang, J. Hagemeister, E. Y. Vedmedenko
and R. Wisendanger, New J. Phys. \textbf{18}, 045021 (2016).

\bibitem{Iwasaki2013} J. Iwasaki, M. Mochizuki, and N. Nagaosa, Nat. Commun.
\textbf{4}, 1463 (2013).

\bibitem{Zhou2014} Y. Zhou and M. Ezawa, Nat. Commun. \textbf{5}, 4652 (2014).

\bibitem{Kai2017} K. Litzius, I. Lemesh, B. Kr\"{u}ger, P. Bassirian, L. Caretta, K. Richter,
F. B\"{u}ttner, K. Sato, O. A. Tretiakov, J. F\"{o}rster, R. M. Reeve1,
M. Weigand, I. Bykova, H. Stoll, G. Sch¨¹tz, G. S. D. Beach,
and M. Kl\"{a}ui, Nat. Phys. \textbf{13}, 170 (2017).

\bibitem{Iwasaki2014} J. Iwasaki, A. J. Beekman, and N. Nagaosa,
Phys. Rev. B \textbf{89}, 064412 (2014).

\bibitem{Schutte2014} C. Schutte and M. Garst,
Phys. Rev. B \textbf{90}, 094423 (2014).

\bibitem{Wang2015} W. Wang, M. Beg, B. Zhang, W. Kuch, and H. Fangohr,
Phys. Rev. B \textbf{92}, 020403 (R) (2015).

\bibitem{Yan2017} W. Yang, H. Yang, Y. Cao, and P. Yan, Optical Express \textbf{26}, 8778 (2018).

\bibitem{Kong2013} L. Kong and J. Zang, Phys. Rev. Lett. \textbf{111}, 067203 (2013).

\bibitem{Lin2014} S.-Z. Lin, C. D. Batista, C. Reichhardt, and A. Saxena,
Phys. Rev. Lett. \textbf{112}, 187203 (2014).

\bibitem{Mochizuki2014} M. Mochizuki, X. Z. Yu, S. Seki, N. Kanazawa, W. Koshibae, J. Zang, M. Mostovoy,
Y. Tokura, and N. Nagaosa, Nat. Mater. \textbf{13}, 241 (2014).

\bibitem{Jiang2017} W. Jiang, X. Zhang, G. Yu, W. Zhang, X. Wang,
M. B. Jungfleisch, J. E. Pearson, X. Cheng, O. Heinonen, K. L. Wang,
Y. Zhou, A. Hoffmann, and S. G. E. te Velthuis, Nat. Phys. \textbf{13}, 162 (2017).

\bibitem{Peng2011} P. Yan, X. S. Wang, and X. R. Wang, Phys. Rev. Lett.
\textbf{107}, 177207 (2011).

\bibitem{Matsukura2015} F. Matsukura, Y. Tokura, and H. Ohno, Nat. Nanotech.
\textbf{10}, 209 (2015) and the references therein.

\bibitem{Ohno2000} H. Ohno, D. Chiba, F. Matsukura, T. Omiya, E. Abe, T. Dietl, Y. Ohno,
and K. Ohtani, Nature \textbf{408}, 944 (2000).

\bibitem{Ando2016} F. Ando, H. Kakizakai, T. Koyama, K. Yamada,
M. Kawaguchi, S. Kim, K.-J. Kim, T. Moriyama, D. Chiba, and T.
Ono, Appl. Phys. Lett. \textbf{109}, 022401 (2016).

\bibitem{Dohi2016} T. Dohi, S. Kanai, A. Okada, F. Matsukura, and H. Ohno,
AIP Advances \textbf{6}, 075017 (2016).

\bibitem{Yang2016} H. Yang, O. Boulle, V. Cros, A. Fert, and M. Chshiev,
arXiv:1603.01847v2.

\bibitem{Weisheit2007} M. Weisheit, S. F\"{a}hler, A. Marty, Y. Souche,
C. Poinsignon, and D. Givord, Science \textbf{315}, 349 (2007).

\bibitem{Maruyama2009} T. Maruyama, Y. Shiota1, T. Nozaki, K. Ohta, N. Toda, M. Mizuguchi, A. A. Tulapurkar, T. Shinjo,
M. Shiraishi, S. Mizukami, Y. Ando, and Y. Suzuki, Nat. Nanotech. \textbf{4}, 158 (2009).

\bibitem{Lebeugle2009} D. Lebeugle, A. Mougin, M. Viret, D. Colson, and L. Ranno,
Phys. Rev. Lett. \textbf{103}, 257601 (2009).

 \bibitem{Heron2011} J. T. Heron, M. Trassin, K. Ashraf, M. Gajek, Q. He, S. Y. Yang,
 D. E. Nikonov, Y. H. Chu, S. Salahuddin, and R. Ramesh, Phys. Rev. Lett. \textbf{107}, 217202 (2011).

\bibitem{Sch2011} A. J. Schellekens, A. van den Brink, J. H. Franken, H. J. M. Swagten,
and B. Koopmans, Nat. Commun. \textbf{3}, 847(2012).

\bibitem{Chiba2012} D. Chiba, M. Kawaguchi, S> Fukami, N. Ishiwata, K. Shimamura,
K. Kobayashi, and T. Ono, Nat. Commun. \textbf{3}, 888 (2012).

\bibitem{Franke2015} K. J. A. Franke, B. VandeWiele, Y. Shirahata, S. J. Hamalainen,
T. Taniyama, and S. vanDijken, Phys. Rev. X \textbf{5}, 011010 (2015).

\bibitem{Dzy1957} I. E. Dzyaloshinskii, Sov. Phys. JETP \textbf{5}, 1259 (1957).

\bibitem{Moriya1960} T. Moriya, Phys. Rev. \textbf{120}, 91 (1960).

\bibitem{Up2015} P. Upadhyaya, G. Yu, P. K. Amiri, and K. L. Wang,
Phys. Rev. B \textbf{92}, 134411 (2015).

\bibitem{mumax3} A. Vansteenkiste, J. Leliaert, M. Dvornik, M. Helsen, F. Garcia-
Sanchez, and F. B. V. Waeyenberge, AIP Adv. \textbf{4}, 107133 (2014).

\bibitem{xiansi2017} X. S. Wang, H. Y. Yuan, and X. R. Wang, Commun. Phys.
\textbf{1}, 31 (2018).

\bibitem{Rohart2013}S. Rohart and A. Thiaville, Phys. Rev. B \textbf{88}, 184422 (2013).

\bibitem{Kravchuk2018} V. P. Kravchuk, D. D. Sheka, U. K. Rossler, J. vandenBrink,
and Y. Gaididei, Phys. Rev. B \textbf{97}, 064403 (2018).




\bibitem{Yin2016} Y. Zhang, X. S. Wang, H. Y. Yuan, S. S. Kang, H. W. Zhang, and X. R. Wang,
J. Phys.: Condens. Matter \textbf{29}, 095806 (2017).

\bibitem{Mochizuki2012} M. Mochizuki, Phys. Rev. Lett. \textbf{108}, 017601 (2012).

\bibitem{Onose2012} Y. Onose, Y. Okamura, S. Seki, S. Ishiwata, and Y. Tokura,
Phys. Rev. Lett. \textbf{109}, 037603 (2012).

\bibitem{Kim2014} J.-V. Kim, F. Garcia-Sanchez, J. Sampaio, C. Moreau-Luchaire, V. Cros, and
A. Fert, Phys. Rev. B \textbf{90}, 064410 (2014).


\bibitem{xiansi2012} X. S. Wang, P. Yan, Y. H. Shen, G. E. W. Bauer, and X. R. Wang,
Phys. Rev. Lett. \textbf{109}, 167209 (2012); X. S. Wang and X. R. Wang, Phys. Rev. B
\textbf{90}, 184415 (2014)



\bibitem{Fasano} A. Fasano and S. Marmi, Analytical Mechanics, Oxford University
Press. (Oxford, New York, 2002).

\bibitem{Schirhagl2014} R. Schirhagl, K. Chang, M. Loretz, and C. L. Degen,
 Annu. Rev. Phys. Chem. \textbf{65}, 83 (2014).

\bibitem{Sakura} J. J. Sakura and J. Napolitano, Modern Quantum Mechanics, 2nd edition,
 (Addison-Wesley, 2011, Boston, Columbus et al.))

%





%
\end{thebibliography}
\end{document}